\documentclass[aps,prl,twocolumn,preprintnumbers,
superscriptaddress,floatfix]{revtex4}

\usepackage{amssymb,multirow}
\usepackage{graphicx}

\begin{document}

\newcommand{\Tr}{\mbox{Tr\,}}
\newcommand{\beq}{\begin{equation}}
\newcommand{\eeq}{\end{equation}}
\newcommand{\bea}{\begin{eqnarray}}
\newcommand{\eea}{\end{eqnarray}}
\renewcommand{\Re}{\mbox{Re}\,}
\renewcommand{\Im}{\mbox{Im}\,}

\voffset 1cm

\newcommand\sect[1]{\emph{#1}---}

\preprint{
\begin{minipage}[t]{3in}
\begin{flushright} SHEP-08-22
\\[30pt]
\hphantom{.}
\end{flushright}
\end{minipage}
}

\title{R-Charge Chemical Potential in a 2+1 Dimensional System}

\author{Nick Evans and Ed Threlfall}

\affiliation{School of Physics and Astronomy, University of
Southampton,
Southampton, SO17 1BJ, UK \\
evans@phys.soton.ac.uk,ejt@phys.soton.ac.uk}

\begin{abstract}
\noindent We study probe D5 branes in D3 brane AdS$_5$ and AdS$_5$-Schwarzschild backgrounds as a prototype dual description of
strongly coupled 2+1 dimensional quasi-particles. We introduce a
chemical potential for a weakly gauged U(1) subgroup of the
theory's global R-symmetry by spinning the D5 branes. The
resulting D5 embeddings are complicated by the existence of a
region of the space in which the local speed of light falls below
the rotation speed. We find regular embeddings through this region
and show that the system does not exhibit the spontaneous symmetry
breaking that would be needed for a superconductor.
\end{abstract}

\maketitle

\section{Introduction}
Recently there has been interest in whether the AdS/CFT
Correspondence \cite{Malda,Witten:1998qj,Gubser:1998bc} can be
used to understand  2+1 dimensional condensed matter systems (for
example
\cite{Hartnoll:2008hs,Hartnoll:2008vx,Roberts:2008ns,Gubser:2008wv,Gubser:2008wz}).
The typical UV degrees of freedom in these systems are electrons
in the presence of a Fermi surface and a gauged U(1), QED. When
brought together in certain 2d states they can become relativistic
and strongly coupled - possibly such systems might induce
superconductivity too by breaking the gauge symmetry. The
philosophy, which may be overly naive, is to find relativistic
strongly coupled systems that show these behaviours and hope they
share some universality with the physical systems. Whether or not
that linkage becomes strong, it is interesting to study the AdS
duals of 2+1d systems.

In this paper we will study the dynamics of the theory on the
world volume of a mixed D3 and D5 brane construction with a 2+1
dimensional intersection$^1$\footnotetext[1]{Mea Culpa: in the
first preprint version of this paper we advertised our computation
as applying to the M2-M5 intersection - this was incorrect since
we had dropped a crucial factor of 2 in the S$^7$ radius. We are
grateful to Veselin Filev for breaking this to us gently! In fact
though the precise zero temperature action studied matches that of
the D3-D5 intersection as we now describe in the text - the
computations can be so easily translated since the form of the
probe D5 action is in this context prescribed by the dimension of
the intersection and the search for flat embeddings at zero
chemical potential - the supersymmetric D3-D5 intersection matches
these conditions. We have updated the thermal computation
although the results are qualitatively the same. We hope to return
to the more complicated M2-M5 system in the future.}, which has
previously been studied at zero temperature in the absence of
chemical potential in
\cite{Karch:2000gx,DeWolfe:2001pq,Erdmenger:2002ex}. The gravity
dual of the D3s, at zero temperature, is $AdS_5 \times S^5$, which
is dual to the 3+1 dimensional ${\cal N}=4$ super Yang Mills
theory. Here these interactions will be used to loosely represent
strongly coupled ``phonons". We will introduce 2+1d ``quasi-particles"
via D5 branes (with a 2+1d intersection with the D3s) - states connecting the two set of branes should be
expected to carry quantum numbers that interact with the D3 brane
dynamics and flavour quantum numbers associated with the number of
D5 branes - the full field theory can be found in
\cite{Erdmenger:2002ex}. We will work in the probe approximation
for the D5 branes which corresponds to quenching quasi-particle
loops in the phonon background \cite{Karch}. At zero chemical
potential the theory has ${\cal N}=4$ supersymmetry and at zero
quasi-particle mass is conformal
\cite{Karch:2000gx,DeWolfe:2001pq,Erdmenger:2002ex}. The system is
related to the higher dimensional D3-D7 intersection where the
${\cal N}=4$ gauge theory on the D3 branes has been used to
describe gluon dynamics and the D3-D7 strings quarks - some
progress in the study of the properties of mesons in 3+1d strongly
coupled gauge theories has been achieved \cite{Erdmenger:2007cm}.
The D3-D5 defect system seems a natural starting point therefore
for 2+1 dimensional systems.

The D3-D5 world volume theory has an unbroken SO(3) global
symmetry. We will imagine gauging an SO(2)/U(1) subgroup of this
to play the role of QED in the solid state system. We will
introduce a chemical potential for the quasi-particles with
respect to the U(1) - this can be done by simply spinning the D5
branes in the SO(2) plane \cite{Albash:2006bs}. The embedding of
the D5 brane is described by a scalar that is charged under this
U(1) symmetry so one naively expects to trigger superconductivity
in the spirit described in \cite{Hartnoll:2008vx} - but here we
would have an explicit understanding of the UV degrees of freedom
the scalar describes. Naively one expects the scalar describing
the D5 embedding to be destabilized by the presence of a chemical
potential which gives the scalar a negative mass squared. We find
the minimum area embedding for such spinning probe D5 branes
though and find this is not the case.

The crucial physics is that the speed of light decreases as one
moves into the centre of AdS - eventually it becomes less than the
rotation speed of the D5 brane. We show, following the higher
dimensional analysis in
\cite{Albash:2006bs,Albash:2007bq,Filev:2008xt} that there are
regular D5 embeddings into the interior which have a more
complicated embedding structure. The branes bend in the direction
of the rotation so that there are two linked scalar fields
describing the embedding - this richer theory turns out to not
include superconductivity, a subtlety on top of the arguments in
\cite{Hartnoll:2008vx}.

We can introduce mass terms for the quasi-particles that
explicitly break the U(1) symmetry and we discuss the embeddings
in these cases. There is a first order phase transition when the
chemical potential grows above the mass of the quasi-particle
bound states - below the transition the quasi-particles exist as
deconfined particles whilst above it they are confined into bound
states. This transition is analogous to the meson melting
transition seen in this system and the D3-D7 system at finite
temperature \cite{Babington,Zamaklar,Hoyos:2006z,Mateos1}. We also
analyze the finite temperature behaviour of these solutions by
using the $AdS_5$ Schwarzschild geometry as the background.

\section{The D3 Theory}

We will represent the strong interaction dynamics with the large N
${\cal N}=4$ super Yang Mills theory on the surface of a stack of
D3 branes. It is described at zero temperature by AdS$_5\times
S^5$ \beq
\begin{array}{ccl}ds^2 & = & {(\rho^2+r^2) \over L^2} dx_{3+1}^2
\\&&\\ &&+ {L^2 \over (\rho^2+r^2)} (d\rho^2 + \rho^2 d \Omega_2^2
+ dr^2 + r^2 d \tilde{\Omega}_2^2) \label{ads4}
\end{array}\eeq where we have written the geometry to display the
directions the D3 lie in ($x_{3+1}$), those we will embed the D5
on ($x_{2+1}$, $\rho$ and $\Omega_2$) and those transverse ($r$
and $\tilde{\Omega}_2$). $L$ is the AdS radius.

At finite temperature the description is given by the
AdS-Schwarzschild black hole \beq ds^2= {u^2 \over L^2} (- h(u)
dt^2 + dx_3^2) + {L^2 \over u^2 h(u)} du^2 + L^2 d \Omega_5^2\eeq
\beq  h(u) = 1 - {u_0^4 \over u^4} \eeq

It is helpful to make the change of variables to isotropic
coordinates \beq {u \; du \over \sqrt{u^4 - u_0^4}} = {dw \over
w} \eeq and choose the integration constant such that if $u_0=0$
the zero-temperature geometry is recovered \beq  2 w^2 =
u^2 + \sqrt{u^4 - u_0^4} \eeq

The metric can now be written as \beq \begin{array}{ccc} ds^2 & =
& \frac{1}{L^2} \left( w^2+\frac{u_0^4}{4w^2}
 \right ) \left ( - \left (
\frac{w^4-\frac{u_0^4}{4}}{w^4+\frac{u_0^4}{4}} \right )^2 dt^2
+dx_3^2 \right )\\&&\\ && +\frac{L^2}{w^2} \left ( d \rho^2 +
\rho^2 d\Omega_2^2 + dr^2 +r^2 d \tilde{\Omega}_2^2 \right )
\end{array}\eeq
with $w^2 = \rho^2 + r^2$, which shares the coordinate structure
of (\ref{ads4}).

\section{Quenched Matter from a D5 Probe At T=0}

We will introduce quenched matter via a probe D5 brane. The
underlying brane configuration is as follows:
\begin{center}\begin{tabular}{ccccccccccc}
& 0 & 1 & 2 & 3 & 4 & 5 & 6 & 7 & 8 & 9  \\
D3 & - & - & - & - & $\bullet$ & $\bullet$ & $\bullet$ &
$\bullet$ & $\bullet$ & $\bullet$  \\
D5 & - & - & - & $\bullet$ & - & - & - & $\bullet$ & $\bullet$ &
$\bullet$
\end{tabular}\end{center}

In polar coordinates the D5 fills the radial direction of AdS$_5$
and is wrapped on a two sphere.

The action for the D5 is just it's world volume \beq S \sim T \int
d^6\xi  \sqrt{- {\rm det} G}  \sim \int d\rho~ \rho^2 \sqrt{1 +
r^{'2}} \eeq where $T$ is the tension and we have dropped angular
factors on the two-sphere.

This is clearly minimized when $r$ is constant so the D5 lies
straight. The value of the constant is the size of the mass gap
for the quasi-particles. We will mainly be interested in the
conformal case where the constant is zero. Note the general large
$\rho$ solution is of the form \beq r = m + {c \over \rho} +..
\eeq Here $m$ is an explicit  mass term for the quasi-particles in
the Lagrangian and $c$ the expectation value for a
bi-quasi-particle operator - note $m$ has dimension one and $c$
dimension two adding to three as required for a Lagrangian term in
2+1d. The solution with non-zero $c$ is not normalizable in pure
AdS$_5$. Note that when $m=c=0$ the theory is conformal. Including
a non-zero $m$ or $c$ breaks the SO(3) symmetry  ie it breaks one
transverse SO(2) symmetry. From this it is apparent that $m$ and
$c$ carry charge under that U(1). Were $c$ to be non-zero when
$m=0$ it would be an order parameter for the spontaneous breaking
of the U(1) symmetry.

\section{Chemical Potential/Spin}

Our theory as yet lacks the relevant perturbation of the Fermi
surface and the U(1) of QED. We will associate the U(1) with a
subgroup of the SO(3) of the $\tilde{\Omega}_2$ - for concreteness
we will use the angle in the $x_7-x_8$ directions.

To include a chemical potential we will spin the D5 brane in the
angular direction $\phi$ of this U(1) with angular speed $\mu$.

The spinning of the D5 branes implies that the quasi-particles see
a chemical potential. This is in fact a little bit of a peculiar
limit since the background D3 theory also has fields, including
scalars, charged under the U(1). We are not allowing that geometry
to backreact to the chemical potential. In fact we had better not
- the pure D3 theory has a moduli space for separating the D3s in
the transverse 6-plane. Were we to set them spinning they would
scatter to infinity since there is no central force to support
rotation. In the theory on the D3 surface there is a run away
Bose-Einstein condensation. We simply wish to switch off this
physics - it is not what we are interested in - so we forbid such
backreaction. The D3 theory is in an unstable state but will
nevertheless provide some strongly coupled interactions for the
quasi-particles that do see the chemical potential.

\subsection{An Overly Naive Ansatz}

We first look for solutions where the D5 embedding has $\phi = \mu
t$ and we will allow the position $r$ (the radial distance in
$x_7-x_8$) to be a function of $\rho$. The action is \beq S \sim
\int d\rho \; \rho^2 \sqrt{(1 + r^{'2})(1- {L^4 \over
(\rho^2+r^2)^2 } r^2 \mu^2)} \eeq Naively one is expecting the
centrifugal force from the spinning to eject the brane from the
axis at all but the end points where the boundary conditions hold
the brane. This would lead to a spontaneous symmetry breaking or
superconducting state.  We will see that this is what this naive
system tries to achieve.

The equation of motion for $r$ as a function of $\rho$ is easily
computed but unrevealing.
At large $\rho$ the solutions tend to the  no-rotation limit $r
\sim m + \frac{c}{\rho}$.

The (pair of) circle(s) in the $(\rho,r)$ plane described by $L^4
\mu^2 r^2=(\rho^2 + r^2)^2$ is clearly a zero of the action so
branes wrapped there provide a solution to the equation of motion.
Anything going within the locus described by the two circles is
moving faster than the local speed of light and is presumably not
physical. This locus is a stationary limit surface - we  call it
the ergosurface below.

There exist ``Karch-Katz" type solutions \cite{Karch} for
D5-branes that do not encounter the stationary limit surface -
these solutions essentially lie flat above everything plotted in
Fig 1. We want to know what happens to those which have a close
encounter with the ergosurface. It turns out that the curves which
minimize the action like to hit the surface at a right angle. They
then kink onto the surface where they can have zero action.

The relation between $m$ and $c$ for  curves impacting on the
stationary limit surface in this way is shown in Fig 1 - the
presence of non-zero $c$ at $m=0$ appears to indicate spontaneous
breaking of the U(1) symmetry ie superconductivity. Note there is
a first order phase transition between the Karch-Katz embeddings
and those hitting the ergosurface - we will discuss this
transition further below.

\begin{centering}
\begin{figure}[h]
\begin{centering}
\includegraphics[width=60mm]{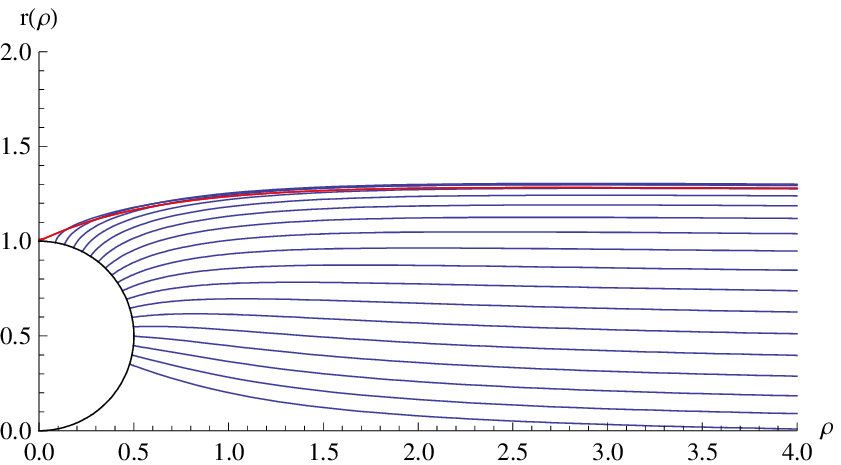} \hspace{1.0cm}
\includegraphics[width=60mm]{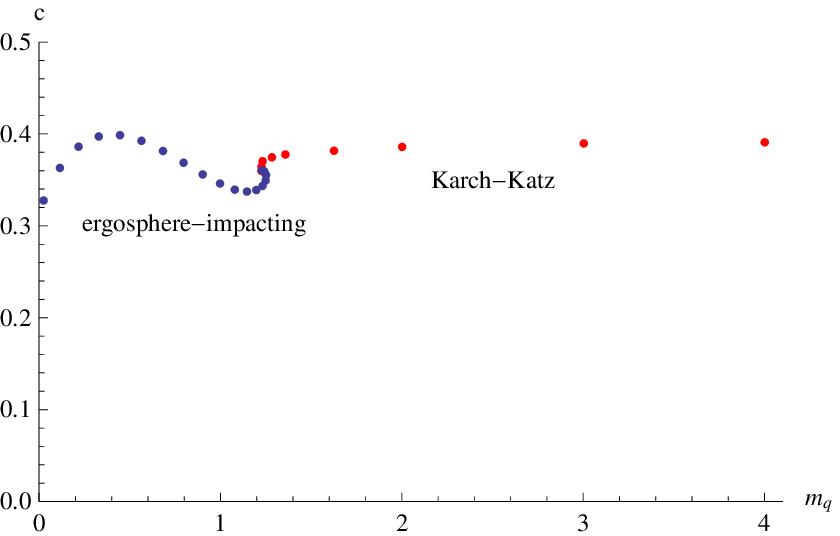}
\par
\end{centering}
\caption{Embeddings of D5 branes impacting on the ergosurface in
blue and the lowest Karch-Katz embedding in red (top). At the
bottom is a plot of $c$ vs $m$ for embeddings (on the left)
impacting on the ergosurface and Karch-Katz embeddings (on the
right). The solutions oscillate around the value for the lowest
Karch-Katz solution as the D5 approaches the very top of the
ergosurface.}
\end{figure}
\par
\end{centering}

The problem of course here is that the solutions are singular at
the ergosurface where they kink. This is a sign that our ansatz is
wrong - none of this is the right physics.

\subsection{A More Sophisticated Ansatz}

We will now try a more sophisticated ansatz where the brane has in
addition some profile $\phi(w )$ where $\phi$ is the angle on
which they spin (ie $\phi = \mu t + \phi(w)$).  The ansatz is
inspired by the work in \cite{Albash:2007bq} where
similar issues are encountered when a magnetic field is switched
on on the brane's world-volume.

We find it numerically convenient to switch coordinates and write the AdS
geometry as
\begin{equation} \begin{array}{ccl}ds^2 & = & \frac{w^2}{L^2} dx_{3+1}^2+\frac{L^2}{w^2}
\left( dw^2 \right. \\ &&\\ && \left. + w^2 \left( d \theta^2 +
\sin^2 \theta d \Omega_2^2 + \cos^2 \theta d \tilde{\Omega}_2^2
\right ) \right ) \end{array}
\end{equation} the D5 will now be embedded in the $x_{2+1}, w$ and
$\Omega_2$ directions - the naive solutions above are recovered by
looking for solutions that have $\theta(w)$ and $\phi = \mu t$
where $\phi$ is the `first' angle of the $\tilde{\Omega}_2$.

In these coordinates the Lagrangian for our more ambitious ansatz
for the rotating D5 embedding is

\begin{equation} \begin{array}{ccl}
\mathcal{L} & = & w^2 \sin^2 \theta \times \\ && \\ && \sqrt{\left
( 1- \frac{L^4 \mu^2 \cos^2 \theta}{w^2} \right )\left ( 1+w^2
\theta'^2 \right )+w^2 \cos^2 \theta \; \phi'^2}
\end{array}
\end{equation}
If $\phi'\sim\mu$ the two $\mu^2$ terms compete against each other
removing the naive intuition about centrifugal force.

Since the action only depends on $\phi'$ and not $\phi$ one can
integrate the equation of motion for $\phi'$. One could then
substitute back in for $\phi'$ in terms of the integration
constant - this though gives an action with a ``zero over zero"
form at the ergosurface that is hard to work with. Instead,
following \cite{Albash:2007bq,Shock:2007}, we Legendre transform to
$\mathcal{L}' \equiv \mathcal{L}- \phi' \frac{\partial
\mathcal{L}}{\partial \phi'}$. This gives (setting $\frac{\partial
\mathcal{L}}{\partial \phi'}=J$)
\begin{equation} \begin{array}{ccc}
\mathcal{L}'&=&\frac{1}{w \cos \theta} \sqrt{\left ( 1-
\frac{L^4 \mu^2 \cos^2 \theta}{w^2} \right )}\sqrt{\left ( 1+w^2
\theta'^2 \right )}\\ &&\\ && \times  \sqrt{\left ( w^6 \sin^4
\theta \cos^2 \theta-J^2 \right )} \end{array}
\end{equation}
This has a ``zero times zero" form at the ergosurface which is
simpler to work with numerically.

For a solution that crosses the ergosurface we demand that the
action be positive everywhere and this fixes $J$ - the two terms
must pass through zero and switch signs together. Having fixed $J$
in this way one can then look at the   $\theta$ equation of motion
near the ergosurface. Expanding near the surface, and after some
algebra, one finds the following consistency equation for the
$\theta$ derivative





\begin{centering}
\begin{figure}[h]
\begin{centering}
\includegraphics[width=80mm]{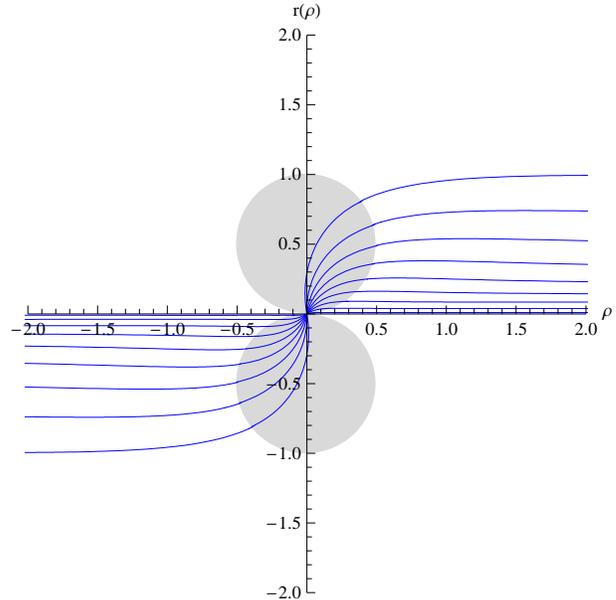}
\par
\end{centering}
\caption{A selection of solution curves for D5 embeddings.  The
grey region is the interior of the ergosurface.}
\end{figure}
\par
\end{centering}

\begin{centering}
\begin{figure}[h]
\begin{centering}
\includegraphics[width=60mm]{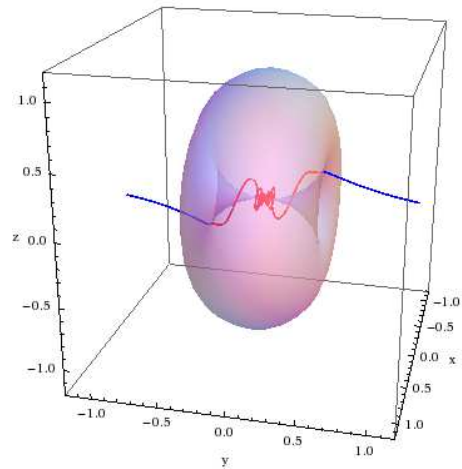}
\par
\end{centering}
\caption{A particular solution curve in the three dimensional
$(w,\theta, \phi)$ subspace. The torus represents the ergosurface.
Note the D5 rotates at speed $\mu$ in the $\phi$ direction (around
the symmetry axis).}
\end{figure}
\par
\end{centering}

\begin{equation}
w^2~ \theta'^2 + \tan \theta~ w ~\theta'-1=0
\end{equation}

There are thus two allowed gradients at the ergosurface. In
fact numerically we find choosing any gradient focuses on to the
same flow both within and outside the ergosurface. We can
numerically shoot in and out from a point near the ergosurface
in order to generate regular embeddings.

In the three-dimensional $(w, \theta, \phi)$ subspace the
ergosurface  is the torus given by $L^2 \mu \cos \theta = \pm w$,
which in a plane of constant $\phi$ gives two adjacent circles of
radius $\frac{\mu L^2}{2}$.  Fig.2 shows a sequence of regular
solutions in the $(\rho, r(\rho))$ coordinates of the previous
section. To obtain regular solutions one should make an odd
continuation to the negative quadrant as shown. We show a full D5
embedding in Fig.3 with both the $\theta(w)$ and $\phi(w)$
dependence plotted - note the D5 rotates at speed $\mu$ in the
$\phi$ direction (around the axis of the torus).

\begin{centering}
\begin{figure}[h]
\begin{centering}
\includegraphics[width=60mm]{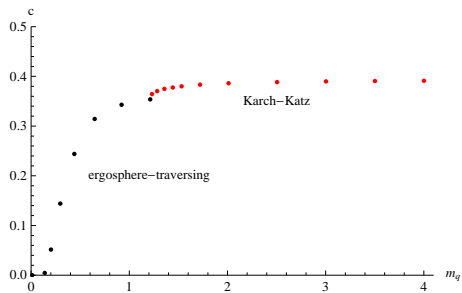}
\par
\end{centering}
\caption{A plot of $c$ vs $m$ for embeddings (on the left)
smoothly penetrating the ergosurface and Karch-Katz embeddings (on
the right). The solutions again oscillate around the value for the
lowest Karch-Katz solution as the D5 approaches the very top of
the ergosurface indicating a first order transition. }
\end{figure}
\par
\end{centering}

Clearly there is no spontaneous symmetry breaking in these
solutions - the solutions smoothly map onto the solution which
lies along the axis as the mass parameter $m$ is taken to zero. In
the field theory presumably the conformal symmetry breaking
parameter ($\mu$) which might trigger symmetry breaking is the
same parameter as that telling us there's a plasma density cutting
off the theory - there's no room for dynamics. This model turns
out not to be an example of the behaviour studied in
\cite{Hartnoll:2008vx}.

The presence of a non-trivial profile $\phi(w)$ for the embeddings
that penetrate the ergosurface indicates on the field theory side
of the duality that there is a vev for the scalar field associated
with the phase of the condensate $c$ - this would be the Goldstone
mode if there were spontaneous symmetry breaking. Note that the
regular Karch Katz embeddings, away from the ergosurface, have
$\phi(w)$ constant so there is no such vev.

Again we see there is a first order transition between the
Karch-Katz type solutions and those that enter the ergosurface
region. We plot the values of $c$ vs $m$ for these solutions in
Fig. 4 - it shows the same spiral structure around the first order
transition as we saw with the naive ansatz. We will discuss the
meaning of this transition below in the thermal context.

\section{Thermal behaviour}

One can perform the same analysis in the thermal background.
Writing $b^4 \equiv \frac{u_0^4}{4}$, there is again a torus-like
ergosurface given by the equation
\begin{equation}
L^2 \mu \cos \theta= \pm \frac{1}{w} \frac{w^4-b^4}{\sqrt{w^4+b^4}}
\end{equation}
and also a spherical horizon at $w=b$.  One finds the horizon
always lies within the ergosurface because the local speed of
light is  zero at the horizon.  Note, below we find no phase
transition when raising the temperature through the scale of the
chemical potential. There would be a transition from a runaway
Bose-Einstein condensation to a stable theory were we to allow the
chemical potential to backreact on the geometry.

One can form the Legendre-transformed Lagrangian (which recovers the
$T=0$ case for $b=0$) \beq \begin{array}{ccc} \mathcal{L}& =&
\frac{1}{w c_{\theta}} \frac{w^4+b^4}{w^4-b^4} \sqrt{\frac{(w^4-b^4)^2}{w^4(w^4+b^4)}\left (1-L^4 \mu^2 c_{\theta}^2 w^2
\frac{(w^4+b^4)}{(w^4-b^4)^2)} \right )} \\&&\\ && \sqrt{ \left ( 1+
w^2 \theta'^2 \right )} \sqrt{ \left ( s_{\theta}^4
c_{\theta}^2 \frac{(w^4-b^4)^2}{w^2} - J^2 \frac{w^4}{w^4+b^4}\right )}
\end{array}\eeq

\begin{centering}
\begin{figure}[h]
\begin{centering}
\includegraphics[width=60mm]{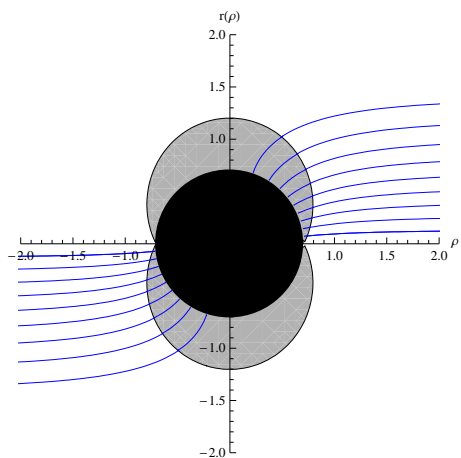}
\par
\end{centering}
\caption{A selection of solution curves for D5 branes in the
thermal geometry (with $u_0=1$).   The grey region is the interior of the
ergosurface and the black region is the interior of the event horizon.}
\end{figure}
\par
\end{centering}

The embeddings which extremize the  action fall into two types -
Karch-Katz type embeddings and those which hit the ergosurface.
Fluctuations of the former would reveal a bound state spectrum.
The latter embeddings inevitably fall onto the event horizon (a
selection of these is plotted in Fig.5 for $u_0=1$). In addition
for these embeddings that pass through the ergosurface $g_{tt}$
switches sign on the world volume - the ergosurface acts like a
horizon for the world volume fields \cite{Filev:2008xt}. Here
fluctuations would have a quasinormal spectrum along the lines of
\cite{Hoyos:2006z}.

There is therefore a first order transition in the behaviour of
the theory as the quasi-particle mass goes through the scale of
the chemical potential or temperature. This transformation is
explored in detail in \cite{Filev:2008xt}. Note here it seems the
transition is always a meson melting transition at finite
temperature.  At zero temperature the transition is driven by
quantum rather than thermal fluctuations and has been described in
terms of a metal-insulator transition in \cite{Karch:2007pd}.

\section{The D3-D7 System}

Much of the above parallels results already found in the D3-D7
system \cite{Albash:2006bs,Filev:2008xt}. That system describes an
${\cal N}=2$ 3+1d gauge theory with fundamental matter
hypermultiplets in the gauge background of ${\cal N}=4$ super-Yang
Mills theory. In \cite{Albash:2006bs} an analysis similar to our
``naive ansatz" was performed suggesting spontaneous symmetry
breaking. Those authors have since refined their analysis in a
related system with a background electric field
\cite{Albash:2007bq} and concluded that if regular embeddings are
insisted upon the symmetry breaking is not present (see also \cite{Shock:2007}). Were they to
update \cite{Albash:2006bs} they would find embeddings analogous
to our D5 embeddings above as they indicate in
\cite{Filev:2008xt}.

\section{Summary}

We have proposed probe D5 branes in D3 brane backgrounds as a
plausible dual for a strongly coupled quasi-particle theory in 2+1
dimensions. We introduced a chemical potential (spin) with respect
to a global U(1) symmetry of the theory and found the resulting
regular D5 embeddings. These embeddings do not display spontaneous
symmetry breaking and, indeed, at zero temperature and zero
intrinsic mass the theory is essentially indifferent to the
chemical potential remaining as a state of conformal
quasi-particles. We do show a first order phase transition in the
massive theory as the quasi-particle mass crosses the value of the
chemical potential - on one side the quasi-particles are confined
on the other they are not.

\vspace{1cm}

\noindent {\bf Acknowledgements:} ET would like to thank STFC for
his studentship funding. We are grateful to Rob Myers and Sean
Hartnoll for discussions at the beginning of this project. We
thank Clifford Johnson, Tameem Albash and Veselin Filev for
guidance on their work.

\end{document}